\documentclass[%
 aip,
 amsmath,amssymb,
 reprint,%
]{revtex4-1}

\usepackage{graphicx}
\usepackage{dcolumn}
\usepackage{bm}

\usepackage[utf8]{inputenc}
\usepackage[T1]{fontenc}
\usepackage{mathptmx}
\usepackage{etoolbox}

\makeatletter
\def\@email#1#2{%
 \endgroup
 \patchcmd{\titleblock@produce}
  {\frontmatter@RRAPformat}
  {\frontmatter@RRAPformat{\produce@RRAP{*#1\href{mailto:#2}{#2}}}\frontmatter@RRAPformat}
  {}{}
}%
\makeatother
\begin{document}

\preprint{AIP/123-QED}

\title{Magnetic transition in B2 Al–Cr–Co alloys}

\author{Haireguli Aihemaiti}
\email{haiaih@kth.se}
\affiliation{Department of Materials Science and Engineering, KTH Royal Institute of Technology, Stockholm SE-10044, Sweden}%
\affiliation{Wallenberg Initiative Materials Science for Sustainability, Department of Materials Science and Engineering, KTH Royal Institute of Technology}

\author{Esmat Dastanpour}%
\email{esmatdh@kth.se}
\affiliation{Department of Materials Science and Engineering, KTH Royal Institute of Technology, Stockholm SE-10044, Sweden}%
\affiliation{Wallenberg Initiative Materials Science for Sustainability, Department of Materials Science and Engineering, KTH Royal Institute of Technology}

\author{Shashank Chaturvedi}%
\affiliation{Department of Materials Science and Engineering, KTH Royal Institute of Technology, Stockholm SE-10044, Sweden}%

\author{Shuo Huang}%
\affiliation{Faculty of Materials Science and Chemistry, China University of Geosciences, Wuhan 430074, China}%

\author{Anders Bergman}%
\affiliation{Department of Physics and Astronomy, Uppsala University, Box 516, SE-75120 Uppsala, Sweden}%

\author{Levente Vitos}%
\affiliation{Department of Materials Science and Engineering, KTH Royal Institute of Technology, Stockholm SE-10044, Sweden}%
\affiliation{Wallenberg Initiative Materials Science for Sustainability, Department of Materials Science and Engineering, KTH Royal Institute of Technology}%
\affiliation{Department of Physics and Astronomy, Uppsala University, Box 516, SE-75120 Uppsala, Sweden}%
\affiliation{Research Institute for Solid State Physics and Optics, Wigner Research Center for Physics, P.O. Box 49, H-1525 Budapest, Hungary}

\begin{abstract}
Using Density Functional Theory (DFT) calculations and Monte-Carlo (MC) simulations, we investigate the recently reported magnetic transition in B2 Al–Cr–Co alloys. The Cr sublattice is alloyed with different amounts of Co in the antiferromagnetic (AFM) B2 AlCr binary alloy and the resulting exchange interactions are analyzed within the Heisenberg Hamiltonian framework. DFT results reveal that at low Co concentrations the system favors the AFM order, while at high Co contents a transition to the ferromagnetic (FM) state is observed. Within the FM stability field, the Curie temperature ($T_{\text{C}}$), obtained within the mean-field approximation, is below $\sim$160 K and decreases with Co concentration. The calculated exchange parameters evolve systematically with Co content, and the trends are consistent with the DFT  total energies. The magnetic configurations obtained from MC simulations follow the DFT results at low Cr levels but predict a spin-glass behavior for alloys containing more than 40 at.\% Co on Cr sublattice. These findings provide a fundamental understanding of how the chemistry-driven changes in exchange interactions affect magnetism in the B2 Al–Cr–Co alloys.
\end{abstract}

\maketitle

\section{\label{sec:level1}Introduction}

Magnetic materials can exhibit a significant change in the magnetic entropy across the magnetic transition. A large magnetic entropy change is necessary for an efficient magnetocaloric effect in the so-called magnetocaloric materials. Since most potential applications are at or near room temperature, the focus is mainly on magnetic materials with a transition temperature around room temperature \cite{smith2012materials, law2023current}. B2-type alloys demonstrating an antiferromagnetic-ferromagnetic (AFM-FM) metamagnetic transition, such as FeRh, are among the most promising magnetocaloric materials. However, the use of critical raw materials, such as Rh, is not desirable for sustainable development. Taking into account these factors in the design of new magnetocaloric materials, a recent work put forward a stable B2 $\mathrm{Al_{50}Mn_{38}Ni_{12}}$ alloy using experimental and theoretical tools \cite{huang2024lattice}. Another study demonstrated the stability of B2 phase in $\mathrm{AlCoCrFeMn_{x}Ni}$ (x = 0.0, 0.04, 0.08, 0.12 and 0.16) alloys \cite{larsen2021magnetic}. These results indicate the possibility of synthesizing alternative medium- and high-entropy alloys with a B2 structure that involve several 3$d$ transition elements.

It was experimentally verified that Co-doped AlMn exhibits a ferromagnetic B2 structure with magnetism primarily attributed to Mn atoms \cite{Dastanpour2025}. Motivated by those findings, recently we carried out a systematic study of 3$d$ transition metals (Cr, Mn, Fe, Co and Ni) alloyed with Al in B2 structure \cite{Aihemaiti2025}. We found that the hypothetical B2 AlCr stabilizes in antiferromagnetic state while B2 AlCo is nonmagnetic (NM). In addition, Co was identified as the best B2 stabilizing magnetic 3$d$ metals. Therefore, one could anticipate that doping Co into AlCr stabilizes the B2 phase and induces a magnetic transition. Indeed, our \textit{ab initio} total energy calculations showed that AlCr exhibits an AFM ground state, but with increasing Co concentration, we observed a transition from AFM to FM ordering and vanishing Curie temperature when Cr is entirely replaced by Co \cite{Aihemaiti2025}.

In the present study, we focus on the chemistry dependence of the magnetic state and magnetic exchange interactions in Co-doped Al–Cr alloys. We determine the interactions between Co–Co, Cr–Cr and Co–Cr as a function of Co concentrations. Based on the total energy results, we expect that varying Co concentration results in a competition between AFM and FM couplings. Using the \textit{ab initio} exchange interactions, we construct the Heisenberg Hamiltonian and carry out Monte-Carlo (MC) simulations to find the magnetic state and magnetic transition temperatures as a function of composition. We show that increasing the Co content changes the character of Cr–Cr interactions which yields a change in the low-temperature magnetic state as well. The rest of the magnetic interactions remain small for the entire compositional range but Co has a strong indirect effect on the Cr–Cr interactions. We speculate that close to the chemistry-driven magnetic transition, the present system could present metatmagnetic transition with a sizable magnetocaloric effect. 

This work is organized as follows. In Sec. II, we discuss the theoretical approaches used to determine the total energies and the mechanism behind the magnetic transition. Results are presented and discussed in Sec. III. Here, we present the relative stability of AFM and FM states and the exchange interactions as a function of Co-content. We introduce results from the MC simulations and assess the magnetic state and temperature-dependent magnetization. Finally, we summarize and conclude our work.

\section{\label{sec:level1}Theoretical methodology}

All calculations are based on Density Functional Theory (DFT). We employed the Exact Muffin Tin Orbitals (EMTO) method \cite{Vitos2001, Vitos2007} in combination with the Coherent Potential Approximation (CPA) \cite{Soven1967, Gyorffy1972, Vitos2001PRL} to compute the total energies and magnetic exchange interactions $J_{ij}$. The exchange-correlation term was considered within the generalized gradient approximation (GGA) using the Perdew–Burke–Ernzerhof (PBE) scheme \cite{Perdew1996, Perdew1992}. The Heisenberg Hamiltonian ($H = -\sum\limits_{i,j} J_{ij}\,\boldsymbol{s}_i\cdot\boldsymbol{s}_j$, where $\boldsymbol{s}_i$ is the normalized local magnetic moment vector at site $i$) was constructed using the \textit{ab initio} exchange interactions calculated in terms of chemical composition. The Heisenberg Hamiltonian was used in Monte-Carlo simulations to reveal the magnetic structure at various temperatures. 

In the EMTO calculations, we set up a 13×13×13 $k$-mesh for the FM state and an 11×11×11 $k$-mesh for the AFM states. The FM state was modeled by the conventional unit cell, while the AFM state by a 2×2×2 supercell. The Monte-Carlo simulations were done using the UppASD code \cite{Eriksson2017}. These simulations were carried out on systems containing 10×10×10 unit cells, with 20,000 MC steps for equilibration, after which an additional 20,000 steps were sampled to obtain thermodynamic averages.

\section{\label{sec:level1}Results and discussion}
\subsection{\label{sec:level2}Equation of state}

\begin{figure}
\centering
\includegraphics[scale=0.35]{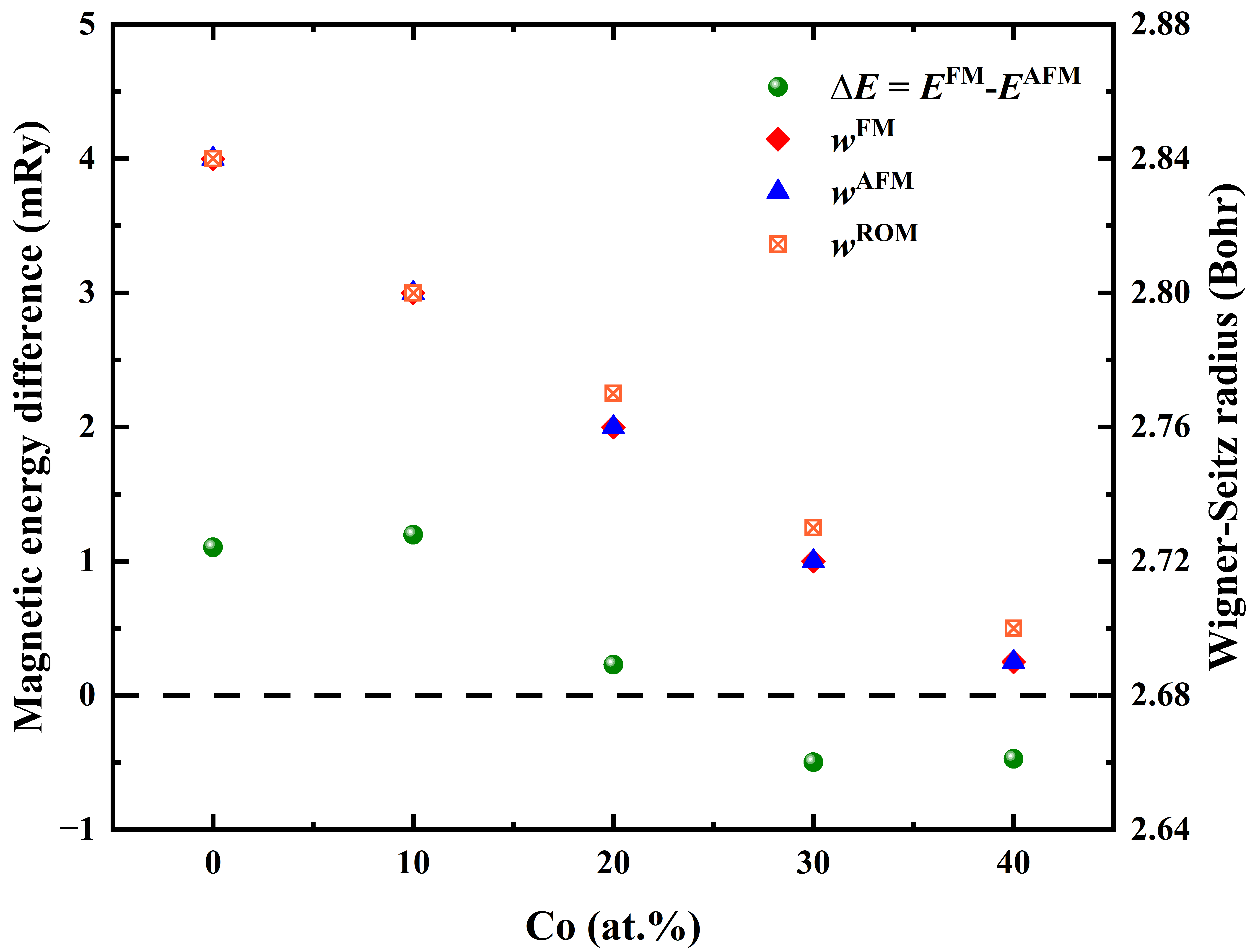}
\caption{\label{fig:1}Calculated magnetic energy difference ($\Delta E$ = $E^{\mathrm{FM}}$ – $E^{\mathrm{AFM}}$) and Wigner-Seitz radii ($w$) for the FM and AFM states for B2 Al–Cr–Co alloys as a function of Co content. ROM stand for the Wigner-Seitz radii estimated from the linear rule of mixture.}
\end{figure}

Before investigating the mechanism of magnetic transition in Al–Cr–Co alloys, we first analyze the total energies of different magnetic states. We performed equation of state (EOS) calculations for these alloys as a function of Co concentration. The total energies correspond to the theoretical equilibrium volumes calculated for each composition. The theoretical equilibrium Wigner-Seitz radii are shown in Fig.~\ref{fig:1} (right axis). They exhibit similar trends for both magnetic structures and follow closely those predicted by the rule of mixture (ROM). Namely, a simple linear interpolation between pure AlCr and AlCo predicts Wigner-Seitz radii 2.80, 2.77, 2.73, and 2.70 Bohr for 10, 20, 30, and 40 at.\% Co, respectively, which are close to the DFT values. The slightly larger DFT values relative to the ROM values for 20, 30 and 40 at.\% Co are due to the magnetic pressure present in the FM and AFM alloys but missing in pure AlCo. 

In Fig.~\ref{fig:1} (left axis), for reference, we also show the total energy difference per atom ($\Delta E$ = $E^{\mathrm{FM}}$ – $E^{\mathrm{AFM}}$) between the FM and AFM states as a function of Co fraction. The present energies are identical with those reported in Ref. \cite{Aihemaiti2025}. The energy difference verifies the possibility of an AFM-FM magnetic transition. Namely, the initially positive magnetic energy difference $\Delta E$ becomes negative for $\mathrm{Al_{50}Cr_{20}Co_{30}}$ and $\mathrm{Al_{50}Cr_{10}Co_{40}}$. According to that, at low Co concentrations, the Al–Cr–Co system is stable in the AFM state, whereas at high Co contents the FM state is more stable than the AFM state. We should notice that based on these total energy results, one cannot rule out the possibility of another magnetic state in intermediate- and high-Co alloys. In order to confirm the stability of the FM state, one needs to carry out a magnetic simulation and find the true low-temperature magnetic state.  

\subsection{\label{sec:level2}Ferromagnetic Al–Cr–Co alloys}

\begin{figure}
\centering
\includegraphics[scale=0.4]{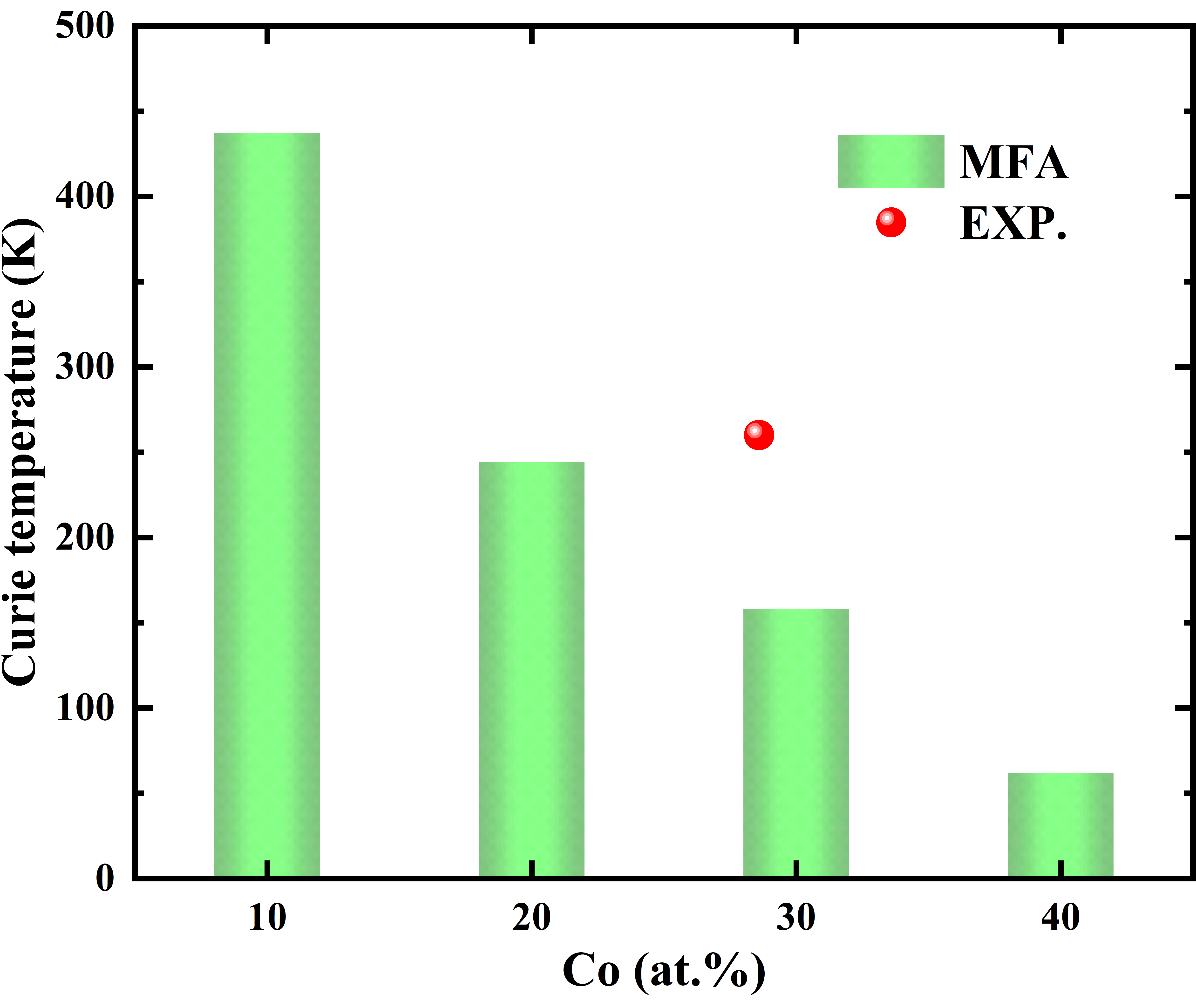}
\caption{\label{fig:4}Theoretical magnetic transition temperatures for the FM state as a function of Co composition calculated using the MFA for the B2 Al–Cr–Co alloys. The red dot represents the experimental $T_{\text{C}}$ value for the arc-melted $\mathrm{Al_{50}Cr_{38}Co_{12}}$ alloy with the B2 phase containing about 28 at.\% Co. \cite{Dastanpour2025b}}
\end{figure}

Figure~\ref{fig:4} shows the Curie temperature of the ferromagnetic state as a function of Co content estimated using the mean-field approximation (MFA). In this approach, $T_{\text{C}}$ is expressed as $T_{\text{C}}$ = $2(E_\mathrm{PM}$ – $E_\mathrm{FM})/3k_\mathrm{B} (1-c)$, where $c$ denotes the concentration of the nonmagnetic component and the paramagnetic (PM) and FM energies are expressed per atom. In the present application, the PM total energies were computed using the static Disordered Local Magnetic moment scheme \cite{Gyorffy1985}.

It is found that Co addition gradually decreases the Curie temperature of the FM state of Al–Cr–Co alloys. We recall that $T_{\text{C}}$ vanishes for AlCo (not shown). The hypothetical FM low-Co alloys have large $T_{\text{C}}$, in line with that predicted by Aihemaiti $et$ $al.$ \cite{Aihemaiti2025} for AlCr. Around 30 at.\% Co, which should be FM according to our DFT total energy calculations (Fig. ~\ref{fig:1}), the estimated Curie point is around 158 K. Interestingly, this MFA value is relatively close to the measured Curie temperature of the Al–Cr–Co ternary alloy containing around 28 at.\% Co  \cite{Dastanpour2025b}. In those experiments, the ternary alloy was prepared by arc melting. Structural analyses demonstrated a two-component system with a non-magnetic Cr-rich body centered cubic phase and a ferromagnetic Co-rich B2 phase. 

\subsection{\label{sec:level2}Exchange interactions}

For spin-dynamics, we need to construct the spin Hamiltonian from the magnetic exchange interactions ($J_{\text{ij}}$). Therefore, next we investigate the Cr–Cr, Co–Cr and Co–Co interactions and to reveal the role these different couplings play. We calculated the exchange interactions at the theoretical equilibrium volumes as a function of composition. We notice that on the simple cubic sublattice occupied by Cr and Co atoms, the first six nearest-neighbor atomic shells are at $a$ (6 atoms), $\sqrt{2}a$ (12 atoms), $\sqrt{3}a$ (8 atoms), $2a$ (6 atoms),  $\sqrt{5}a$ (24 atoms) and $\sqrt{6}a$ (24 atoms), where $a$ is the cubic lattice constant. All exchange interaction calculations were done for the FM state. The interactions between $3d$ metals and Al are very small and thus they are not discussed here.

\begin{figure}
\centering
\includegraphics[scale=0.4]{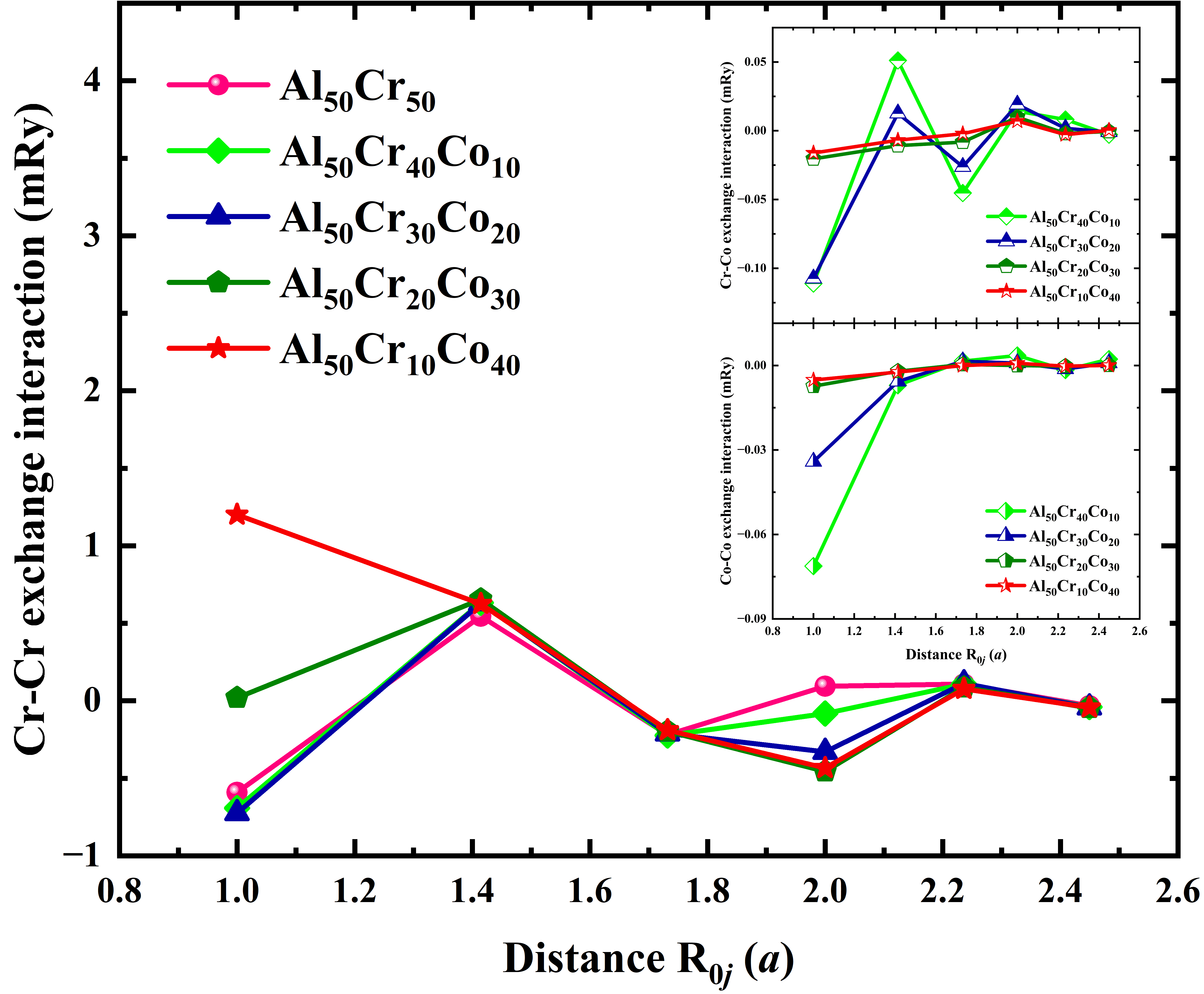}
\caption{\label{fig:2}The calculated Cr–Cr exchange interaction $J_{\text{0j}}$ for B2 AlCr, $\mathrm{Al_{50}Cr_{40}Co_{10}}$, $\mathrm{Al_{50}Cr_{30}Co_{20}}$, $\mathrm{Al_{50}Cr_{20}Co_{30}}$, and $\mathrm{Al_{50}Cr_{10}Co_{40}}$. $\mathrm{R}_{\mathrm{0}j}$ represents the distance between atom $j$ and atom at the origin in units of B2 lattice parameter $a$. The insets show the Cr–Co and Co–Co exchange interactions.}
\end{figure}

Figure \ref{fig:2} displays the calculated Cr–Cr exchange interactions in B2 Al–Cr–Co alloys as a function of Co concentration between 0 and 40 at.\%.  In low-Co containing alloys, such as AlCr, $\mathrm{Al_{50}Cr_{40}Co_{10}}$, and $\mathrm{Al_{50}Cr_{30}Co_{20}}$, the first nearest-neighbor Cr–Cr interaction is negative while the second nearest-neighbor Cr–Cr interaction is positive, indicating antiferromagnetic coupling. This trend is in line with the previously reported AFM state for the B2 AlCr system \cite{Aihemaiti2025}. However, with increasing Co content, the trend changes. In $\mathrm{Al_{50}Cr_{20}Co_{30}}$ and $\mathrm{Al_{50}Cr_{10}Co_{40}}$ alloys, the first nearest-neighbor Cr–Cr interaction changes sign from negative to positive, showing a ferromagnetic coupling between Cr atoms. This change in the sign of the leading Cr–Cr $J_{\text{ij}}$ indicates a magnetic transition driven by chemistry. That is, the leading magnetic interaction changes from antiferromagnetic to ferromagnetic coupling with increasing Co content. We observe another important change in Cr–Cr $J_{\text{ij}}s$. Namely, the originally positive fourth interactions (at distance $2a$) also change sign, and become negative with increasing Co content. The rest of the magnetic interactions are only weakly affected by the chemistry. 

The insets of Fig.~\ref{fig:2} show the calculated Co–Cr and Co–Co magnetic interactions at different Co concentrations. Their $J_{\text{ij}}$ values are much smaller than those of the Cr–Cr interaction. Notice the scale difference between the main panel and the two insets. This finding indicates that magnetism should primarily be driven by the Cr–Cr interaction. This dominance of the Cr–Cr coupling should account for the magnetic transition anticipated in our total energy calculations. We will verify this effect using Monte-Carlo simulations.

\subsection{\label{sec:level2}Magnetic simulations}

Using the above Cr–Cr, Cr–Co and Co–Co magnetic exchange interactions, we carried out MC simulations for alloys containing 10, 20, 30 and 40 at.\% Co. These magnetic simulations as a function of Co concentration allow us to directly assess the role of Co doping in stabilizing ferromagnetism or any other magnetic state in B2 Al–Cr–Co alloys. We should emphasize that in these MC simulations, the Co–Cr and Co–Co interactions are taken into account, but the very weak interactions between magnetic elements and Al (not shown) are omitted. 

\begin{figure}
\centering
\includegraphics[scale=0.39]{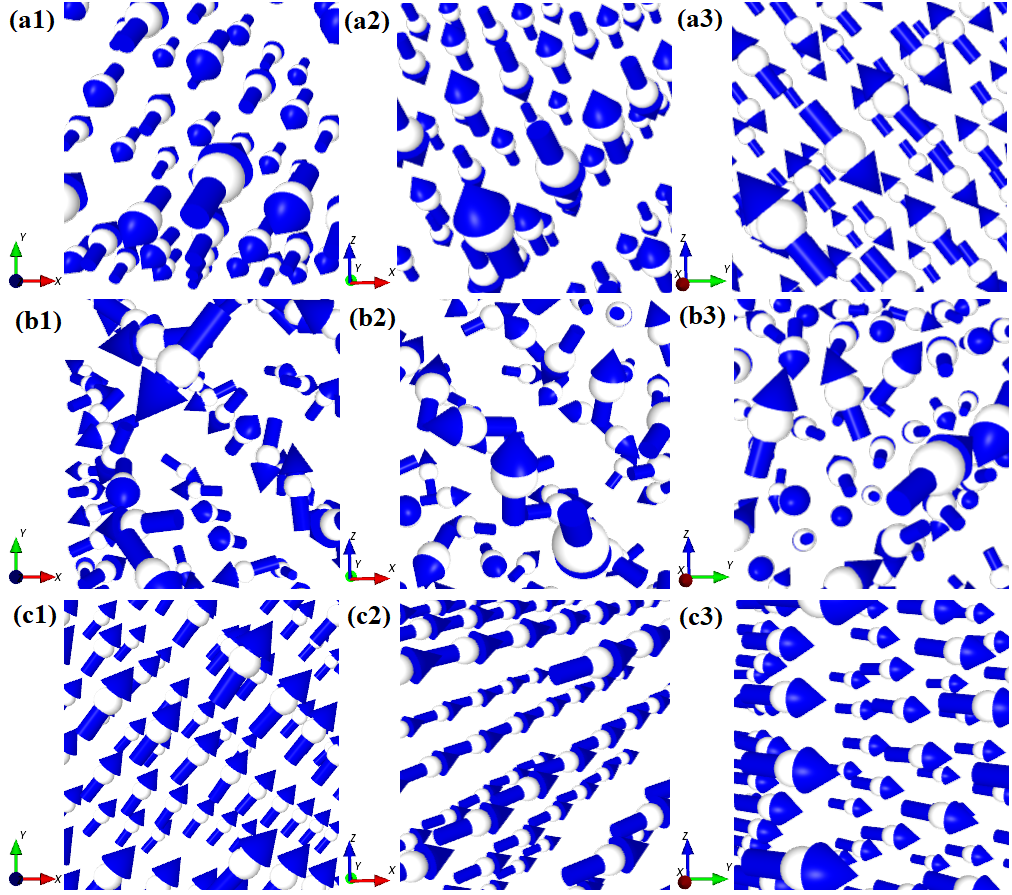}
\caption{\label{fig:3} Illustration of the low-temperature (1 K) antiferromagnetic order found for $\mathrm{Al_{50}Cr_{40}Co_{10}}$ alloy (panels a1-a3), and the disordered magnetic state found for $\mathrm{Al_{50}Cr_{20}Co_{30}}$ (panels b1-b3) in B2 structure. All magnetic configurations are shown in $x$-$y$ plane, $x$-$z$ plane, and $y$-$z$ plane. For comparison, we also show the magnetic state found for the hypothetical $\mathrm{Al_{50}Cr_{50}}$ alloy using the magnetic exchange interactions calculated for  $\mathrm{Al_{50}Cr_{20}Co_{30}}$ (panels c1-c3). Displayed are the randomly distributed Cr and Co magnetic moments in panels a1-b3 and only Cr moments in panels c1-c3.}
\end{figure}

The obtained magnetic configurations at temperature of 1 K are shown in Fig.~\ref{fig:3} for $\mathrm{Al_{50}Cr_{40}Co_{10}}$ and $\mathrm{Al_{50}Cr_{20}Co_{30}}$. At 10 at.\% Co concentration (panels a1-a3), the magnetization vanishes even at low temperature (1 K), and the spins have a preferential magnetic orientation consistent with the antiferromagnetic structure reported in the previous study on AlCr \cite{Aihemaiti2025}. However, when the Co concentration reaches 60 at.\% on the Cr sublattice, the low-temperature state becomes disordered, with Cr and Co magnetic moments having random orientations (Fig.~\ref{fig:3}, panels b1-b3). Therefore, the MC simulations predict that the low temperature magnetic state in Al–Cr–Co alloys undergoes a transition from ordered AFM to a disordered magnetic state as the Co concentration increases. According to the total energy results (Fig. ~\ref{fig:1}), at low temperature (static conditions) the AFM-FM magnetic transition occurs around 22 at.\% Co. However, our MC simulations predict disordered magnetic state already at 20 at.\% Co (not shown). 

It is interesting to investigate in more details the character of the magnetic interactions obtained for 30 at.\% Co. The small but positive nearest-neighbor interaction (0.02 mRy) is followed by a large ferromagnetic coupling (0.65 mRy) and then two sizable antiferromagnetic terms (-0.20 mRy at $\sqrt{3}a$ and -0.45 mRy at 2$a$). The fifth and sixth interactions are 0.09 mRy and -0.05 mRy, respectively. In a simple cubic lattice, assuming 100\% occupancy with Cr, such interactions lead to FM order. This is illustrated in Fig.~\ref{fig:3}, panels c1-c3. Here, all moments represent Cr and they form a clear long-range ferromagnetic order. That is, if all sites were fully occupied by Cr, the Heisenberg Hamiltonian based on the present $J_{\text{ij}}$s obtained for $\mathrm{Al_{50}Cr_{20}Co_{30}}$ would lead to FM order. However, when the simple cubic sites are only partially occupied by Cr and the rest by Co (showing only weak magnetic interactions) like in $\mathrm{Al_{50}Cr_{20}Co_{30}}$, then according to our MC simulations, the long-range order does not form and the system should have a disordered spin-glass type of low-temperature magnetic state. This finding is in contrast to our prediction based on the DFT energies and also to the recent experiments showing a weak ferromagnetic order in the B2 component of the as-cast $\mathrm{Al_{50}Cr_{38}Co_{12}}$ alloy\cite{Dastanpour2025b}. Although the reason for this discrepancy is not yet known, we ascribe it to the exchange interactions computed using the single-site coherent potential approximation in combination with the EMTO method. 

\subsection{\label{sec:level2}The effect of Co on Cr-Cr $J_{\text{ij}}$}

The magnetic simulation results presented in the previous section were obtained using the Cr–Cr, Cr–Co and Co–Co magnetic exchange interactions. These interactions were computed for the actual compositions taking into account the small magnetic moments on Co and Al atoms. Since the moments on Al atoms are around $\sim$0.0 $\mu_B$ and those on Co atoms are $\sim$0.2 $\mu_B$ (the exact value slightly depends on alloy composition), here we consider Co as the most significant medium affecting the Cr–Cr interactions. We should add that neglecting the Cr–Co and Co–Co interactions in the MC simulations yields very similar configurations to those displayed in Fig.~\ref{fig:3}.

To clarify whether Co atoms contribute to the actual magnetic order/disorder, we performed additional $J_{\text{ij}}$ calculations in which the Co magnetic moments were constrained to zero. Here we discuss the results obtained for  $\mathrm{Al_{50}Cr_{20}Co_{30}}$ only. We found that the first nearest-neighbor Cr–Cr interaction becomes negative when the Co moments are fixed to zero. This indicates that although Co–Co and Co–Cr exchange interaction parameters are very small compared to the Cr–Cr interactions, the presence of Co local moments still have an indirect influence on the magnetic interaction between Cr atoms. That is, the weakly magnetic Co atoms mediate the Cr–Cr coupling and turn them FM. When Co carries a finite moment, it alters the energy cost of spin flips on nearby Cr sites, thereby turning the nearest-neighbor exchange interactions ferromagnetic. In summary, at low Co content, Cr–Cr interactions are strongly AFM. Beyond $\sim$30 at.\% Co, the Co atoms mediate a change in the nearest-neighbor Cr–Cr coupling from AFM to FM. The rest of the interactions are much less affected by the Co local magnetic moments. 

\section{\label{sec:level1}Conclusion}

We investigated the magnetic transition in B2 Al–Cr–Co alloys by combining Density Functional Theory calculations, exchange interaction analysis, and Monte-Carlo simulations. Our DFT results show that B2 AlCr stabilizes in an antiferromagnetic ground state, while increasing Co concentration in B2 Al–Cr–Co drives a magnetic transition from antiferromagnetic state to ferromagnetic state. The Curie temperature estimated from the DFT total energies in combination with the mean-field approximation yield a decreasing trend with increasing Co content. For $\mathrm{Al_{50}Cr_{20}Co_{30}}$ alloy, the mean-field critical temperature is in line with the recent experimental findings. The total energy calculations and $J_{\text{ij}}$ trends consistently reveal that above 30 at.\% Co, the dominant magnetic exchange interactions between Cr atoms change sign from antiferromagnetic to ferromagnetic. The MC simulations based on \textit{ab initio} exchange interactions, on the other hand, predict a low-temperature antiferromagnetic state for 10 at.\% Co and a disordered magnetic state above 20 at.\% Co. This prediction is in contrast to the DFT results and recent experiments, calling for further experimental and theoretical works. 

The present findings give a quantitative understanding of how the DFT phase stability and magnetic exchange interactions evolve with Co composition in Al–Cr–Co alloy system. They provide a broader insight into the design of potential new types of B2-type magnetocaloric materials.

\section*{Acknowledgments}

This work was financially supported by the Wallenberg Initiative Materials Science for Sustainability (WISE) funded by the Knut and Alice Wallenberg Foundation (KAW). We also acknowledge the Swedish Foundation for Strategic Research, the Carl Tryggers Foundation, eSSENCE, Olle Engkvists Foundation No. 236-0611 and Formas – a Swedish Research Council for Sustainable Development. The computations were enabled by resources provided by the National Academic Infrastructure for Supercomputing in Sweden (NAISS) at the National Supercomputing Centre (NSC, Tetralith cluster), partially funded by the Swedish Research Council through Grant Agreement No. 2022-06725.

\section*{AUTHOR DECLARATIONS}
\subsection*{Conflict of Interest}
The authors have no conflicts to disclose.

\subsection*{Author Contributions}
\textbf{H. Aihemaiti:} Conceptualization, Investigation, Formal Analysis,
Visualization, Methodology, Writing/Original Draft Preparation,
Writing/Review \& Editing. \textbf{E. Dastanpour:} Formal Analysis, Investigation, 
Writing/Review \& Editing. \textbf{S. Chaturvedi:} Methodology, 
Writing/Review \& Editing. \textbf{S. Huang:} Investigation, Formal Analysis,
Visualization, Methodology, 
Writing/Review \& Editing. \textbf{A. Bergman:} Conceptualization, Investigation, Formal Analysis,
Visualization, Methodology, 
Writing/Review \& Editing.
\textbf{L. Vitos:} Conceptualization, Investigation, Formal Analysis,
Visualization, Methodology, Writing/Original Draft Preparation,
Writing/Review \& Editing.

\section*{References}

\nocite{*}
\bibliography{aipsamp}

\end{document}